\documentclass[
 reprint,
 amsmath,amssymb,
 prl,
 superscriptaddress,
]{revtex4-1}
\usepackage{graphicx}% Include figure files
\usepackage{dcolumn}% Align table columns on decimal point
\usepackage{bm}% bold math
\usepackage{hyperref}% add hypertext capabilities
\usepackage{subfigure}
\usepackage{cleveref}

\newcommand{\subfigimg}[3][,]{%
  \setbox1=\hbox{\includegraphics[#1]{#3}}% Store image in box
  \leavevmode\rlap{\usebox1}% Print image
  \rlap{\hspace*{-3pt}\raisebox{\dimexpr\ht1-0.8\baselineskip}{#2}}% Print label
  \phantom{\usebox1}% Insert appropriate spcing
}

\newcommand{\subfigimgt}[3][,]{%
  \setbox1=\hbox{\includegraphics[#1]{#3}}% Store image in box
  \leavevmode\rlap{\usebox1}% Print image
  \rlap{\hspace*{-7pt}\raisebox{\dimexpr\ht1-0.6\baselineskip}{#2}}% Print label
  \phantom{\usebox1}% Insert appropriate spcing
}
\usepackage{hyperref}
\hypersetup{
 colorlinks=true,
 linkcolor=blue,
 filecolor=magenta,
 urlcolor=cyan,
}

\begin{document}
\preprint{APS/123-QED}
\title{Artificial gauge fields and topological insulators in Moir\'e superlattices}
%\author{{ Ce Shang, Adel Abbout, Xiaoning Zang, Udo Schwingenschl\"{o}gl and Aur\'elien Manchon}\\
%{\small \em King Abdullah University of Science and Technology (KAUST), Physical Science and %Engineering Division,
%Thuwal 23955-6900, Saudi Arabia}}
\affiliation{King Abdullah University of Science and Technology (KAUST), Physical Science and Engineering Division (PSE), Thuwal 23955-6900, Saudi Arabia.}
\affiliation{King Abdullah University of Science and Technology (KAUST), Computer, Electrical and Mathematical Science and Engineering Division (CEMSE), Thuwal 23955-6900, Saudi Arabia.}
\affiliation{King Abdullah University of Science and Technology (KAUST), Physical Science and Engineering Division (PSE), Thuwal 23955-6900, Saudi Arabia.}
\affiliation{Physics Department, King Fahd University of Petroleum and Minerals,
Dhahran 31261, Saudi Arabia}
\author{Ce Shang}
\email{shang.ce@kaust.edu.sa}
\affiliation{King Abdullah University of Science and Technology (KAUST), Physical Science and Engineering Division (PSE), Thuwal 23955-6900, Saudi Arabia.}
\author{Adel Abbout}
\affiliation{King Abdullah University of Science and Technology (KAUST), Physical Science and Engineering Division (PSE), Thuwal 23955-6900, Saudi Arabia.}
\affiliation{Physics Department, King Fahd University of Petroleum and Minerals,
Dhahran 31261, Saudi Arabia}
\author{Xiaoning Zang}
\affiliation{King Abdullah University of Science and Technology (KAUST), Physical Science and Engineering Division (PSE), Thuwal 23955-6900, Saudi Arabia.}
\author{Udo Schwingenschl\"{o}gl}
\affiliation{King Abdullah University of Science and Technology (KAUST), Physical Science and Engineering Division (PSE), Thuwal 23955-6900, Saudi Arabia.}
\author{Aur\'elien Manchon}
\email{aurelien.manchon@kaust.edu.sa}
\affiliation{King Abdullah University of Science and Technology (KAUST), Physical Science and Engineering Division (PSE), Thuwal 23955-6900, Saudi Arabia.}
\affiliation{King Abdullah University of Science and Technology (KAUST), Computer, Electrical and Mathematical Science and Engineering Division (CEMSE), Thuwal 23955-6900, Saudi Arabia.}
%%%%%%%%%%%%%%%%%%%%%

\date{\today}

\begin{abstract}
We propose an innovative quantum emulator based on Moir\'e superlattices  showing that, by employing periodical modulation on each lattice site, one can create tunable, artificial gauge fields with imprinting Peierls phases on the hopping parameters and realize an analog of novel Haldane-like phase. As an application, we provide a methodology to directly quantify the topological invariant in such a system from a dynamical quench process. This design shows a robustly integrated platform which opens a new door to investigate topological physics.
\end{abstract}

\maketitle
{\em Introduction.}---Moir\'e superlattices (MSs), originating from rotational alignment or/and lattice constants mismatch between  individual layers, open up new strategies for engineering electronic properties \cite{PhysRevLett.99.256802,bistritzer2011moire,PhysRevB.81.161405}. When the twist angle is such that the two monolayers become commensurate, their Moir\'e pattern naturally endows (at least approximately) a triangular superlattice \cite{PhysRevB.86.155449}. The size of the Moir\'e unit cell can be parametrized as a periodic function of the twist angle setting out a new length scale orders of magnitude larger than that of the underlying atomic lattices. Consequently, the electrons effectively \lq forget\rq\ the underlying (small) lattice and travel under the effect of the larger Moir\'e lattice. Referring to the approach as \textit{twistronics}, the recent discovery of correlated Mott insulating states and superconducting phases in the twisted bilayer graphene (TBG) \cite {cao2018correlated,cao2018unconventional} has drawn enormous attention and aroused intensive theoretical research \cite{PhysRevX.8.031089,PhysRevX.8.031088,PhysRevX.8.031087,PhysRevLett.121.257001,PhysRevLett.122.106405}.

Gauge fields unveil one of the most ubiquitous concepts which describe many branches of physics, ranging from the standard model \cite{aitchison2012gauge} to the general theory of relativity \cite{zinn1996quantum}. While gauge potentials naturally originate from classical electromagnetism, artificial gauge fields can also be created and tuned by engineering quantum states of matter. In recent years, emulating gauge theories with minimally invasive means has been realized in a plethora of platforms, including  solids \cite{vozmediano2010gauge}, ultracold atoms \cite{PhysRevLett.111.185301,PhysRevLett.111.185302,jotzu2014experimental}, photonics \cite{lumer2019light}, acoustics \cite{xiao2015synthetic}, and mechanical systems \cite{susstrunk2015observation}. In particular,  engineered systems with artificial gauge field \cite{RevModPhys.83.1523,goldman2014light} can also extend their proven quantum simulation abilities further, e.g., to quantum Hall physics or topological insulators \cite{RevModPhys.91.015006,RevModPhys.91.015005}.
Demonstrated to date, lattice modulation techniques have been considered as an efficient means for generating nontrivial topological order and gauge structures in quantum-matter systems \cite{PhysRevX.4.031027,RevModPhys.89.011004,RevModPhys.91.015005}.
Inspired by such “lattice engineering”, we propose to use external field-atom interaction to generate artificial gauge potentials acting on MSs.

In this paper, we introduce and analyze a novel scheme of time-periodic forcing Moir\'e superlattices for facilitating the implementation of artificial gauge fields and investigate the topological properties in such system. By the temporal modulation of the onsite energies, one can induce phases on the tunneling between the Moir\'e sites. Thus the design approach turns MSs into a robust platform with which a host of topological behaviors can be realized and explored. In this frame, we obtain an effective tight-binding model on the emergent MSs according to the Floquet theory, and in which by performing an energy offset between neighbouring sites, an analog of an exotic Haldane-like phase can be realized. To directly identify the topological properties of the system, we adopt a method to extract the Hopf number from the dynamical process after an initial topologically trivial state quench to a final topologically non-trivial one. This quantized value is exactly the same as the Chern number of the post-quench Hamiltonian \cite{PhysRevLett.118.185701}. Taken together, this scheme may open the path to novel device types, unexplored techniques for transport of energy and information, dynamical tools for controlling and probing physical properties, and enlarge the range of systems for constructing physical quantum emulators in Richard Feynman's vision \cite{feynman1982simulating}. Compared with laboratory optical lattices, MSs show more flexibility for integration by reducing component size and complexity.

\begin{figure*}[t!]
\begin{minipage}{0.31\linewidth}
\subfigure{\label{fig:1a}\subfigimg[width=5.5cm,height=6cm]{(a)}{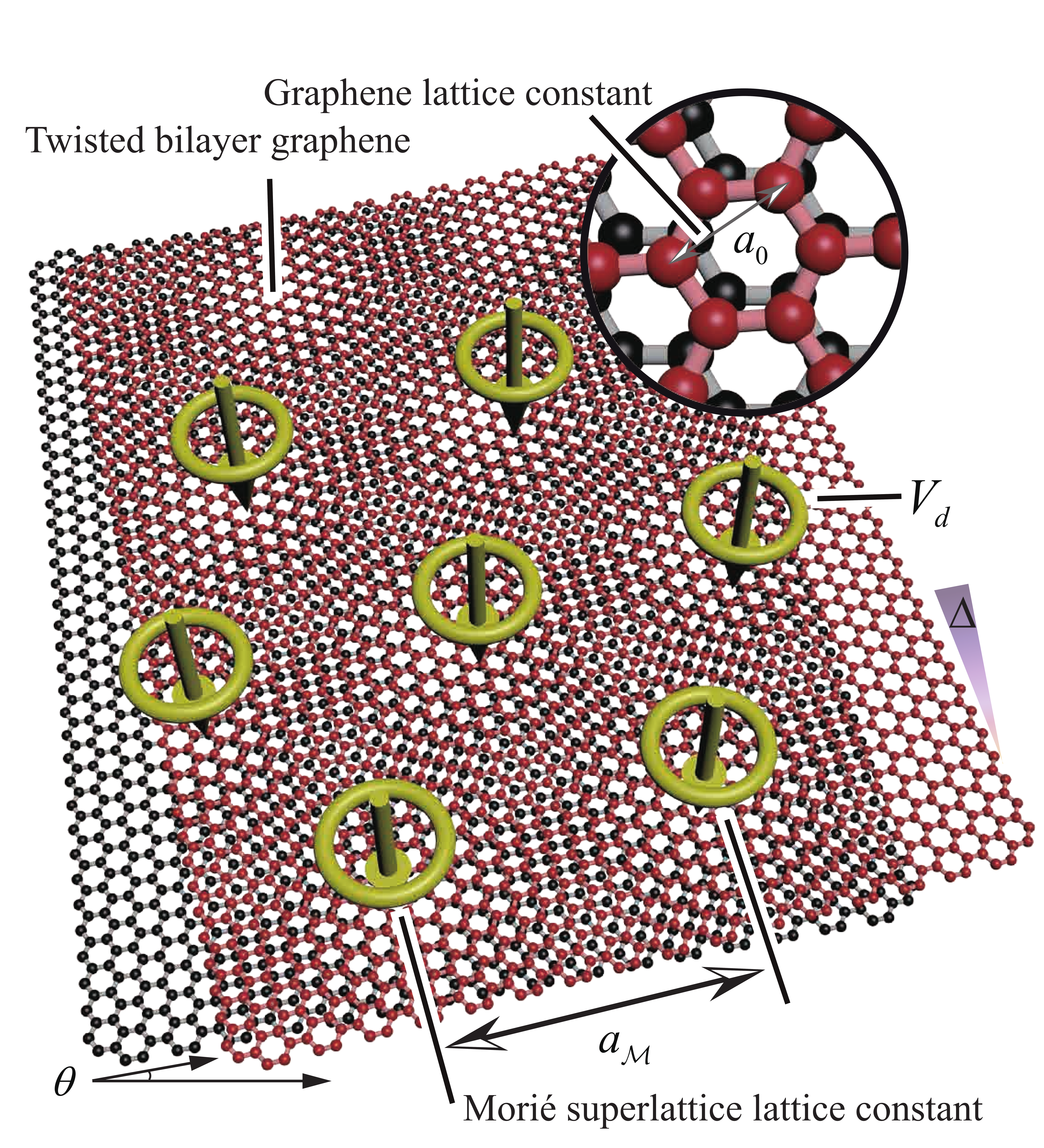}}
\end{minipage}%
\begin{minipage}{0.69\linewidth}
\subfigure{\label{fig:1b}\subfigimg[width=8cm,height=3cm]{(b)}{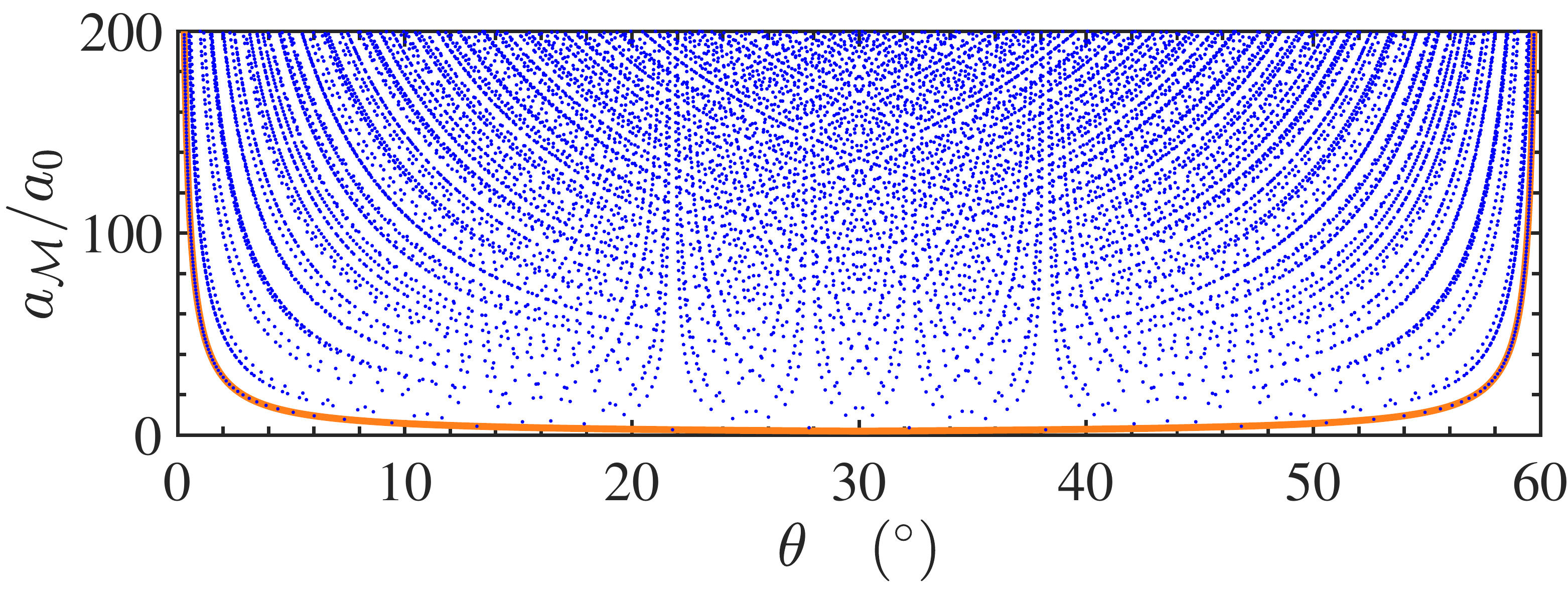}}
\subfigure{\label{fig:1c}\subfigimg[width=3cm,height=3cm]{(c)}{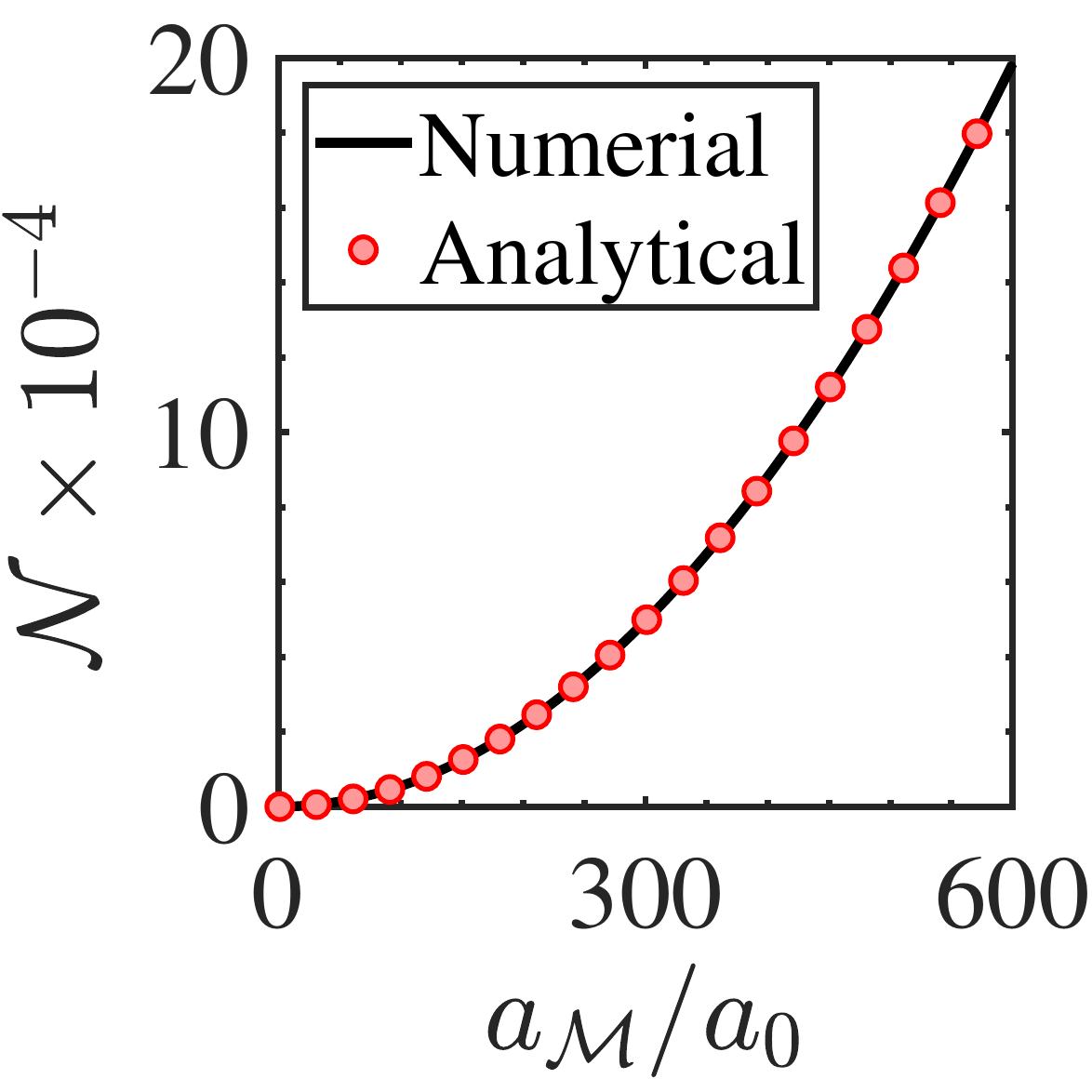}}\\
\subfigure{\label{fig:1d}\subfigimg[width=3.6cm,height=3cm]{(d)}{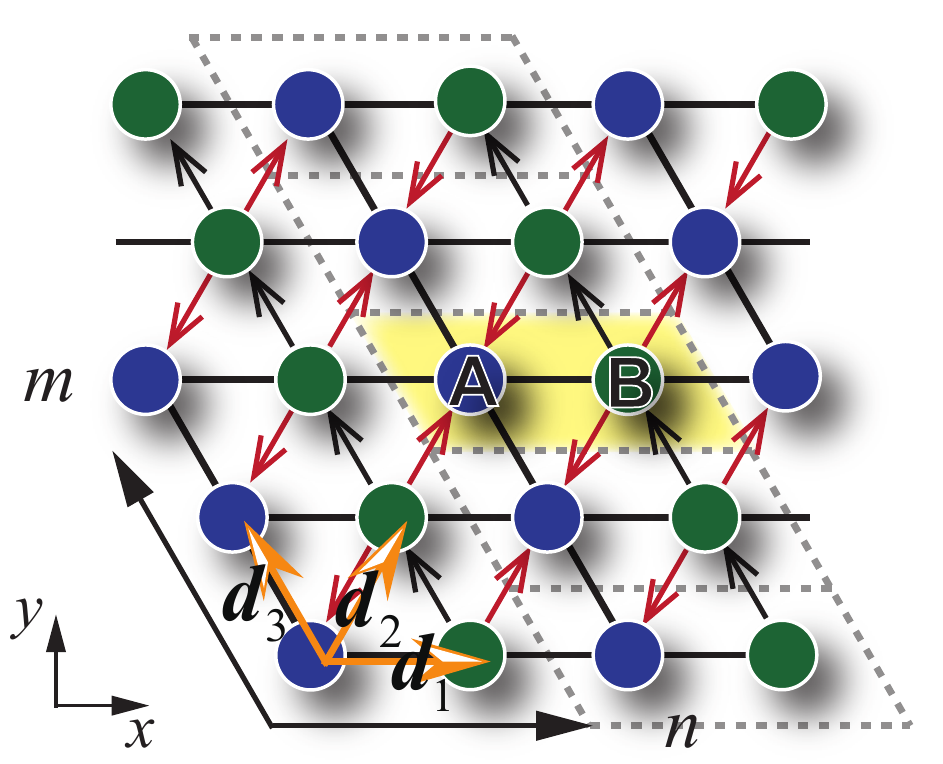}}
\subfigure{\label{fig:1e}\subfigimg[width=3.6cm,height=3cm]{(e)}{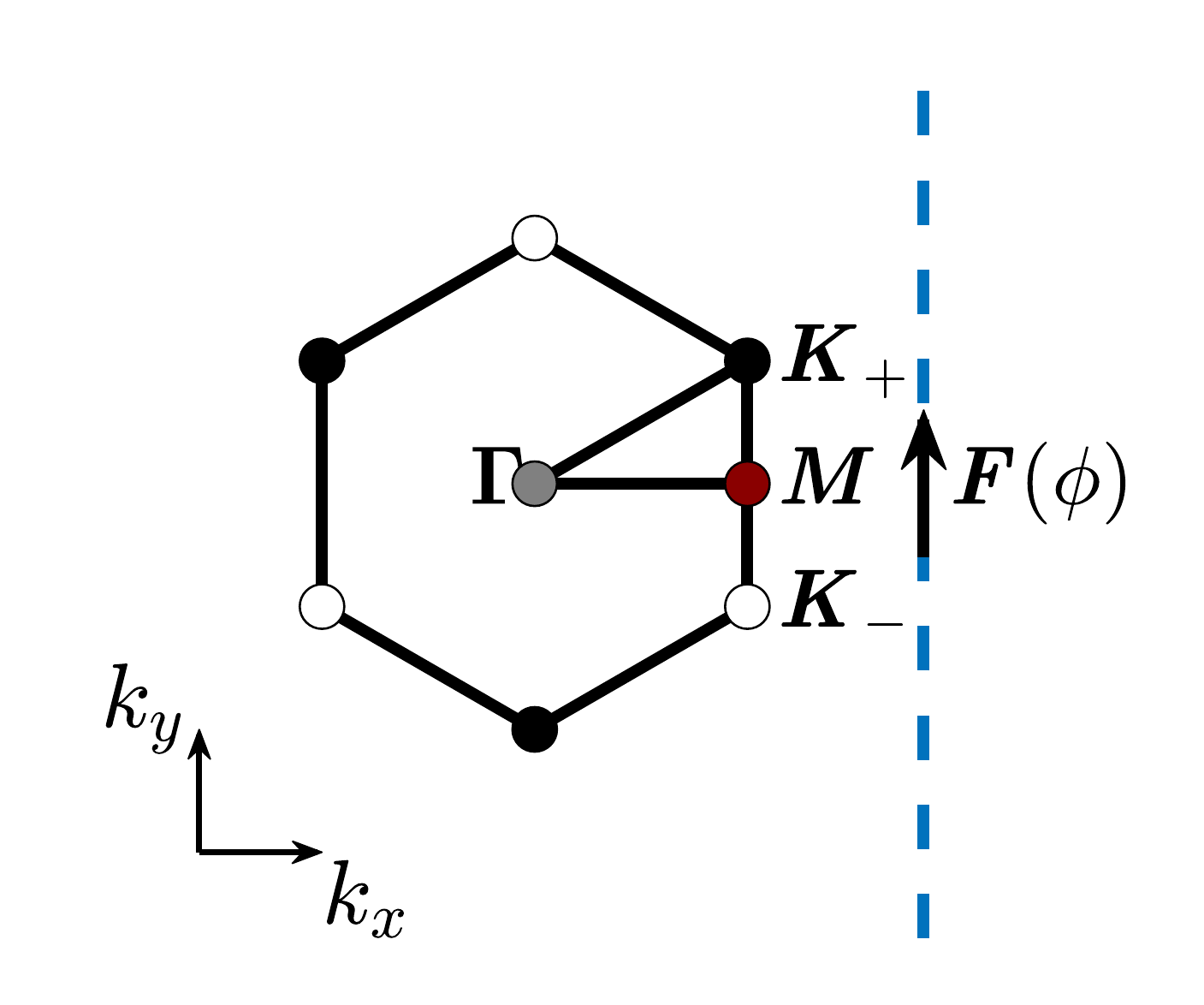}}
\subfigure{\label{fig:1f}\subfigimg[width=3.6cm,height=3cm]{(f)}{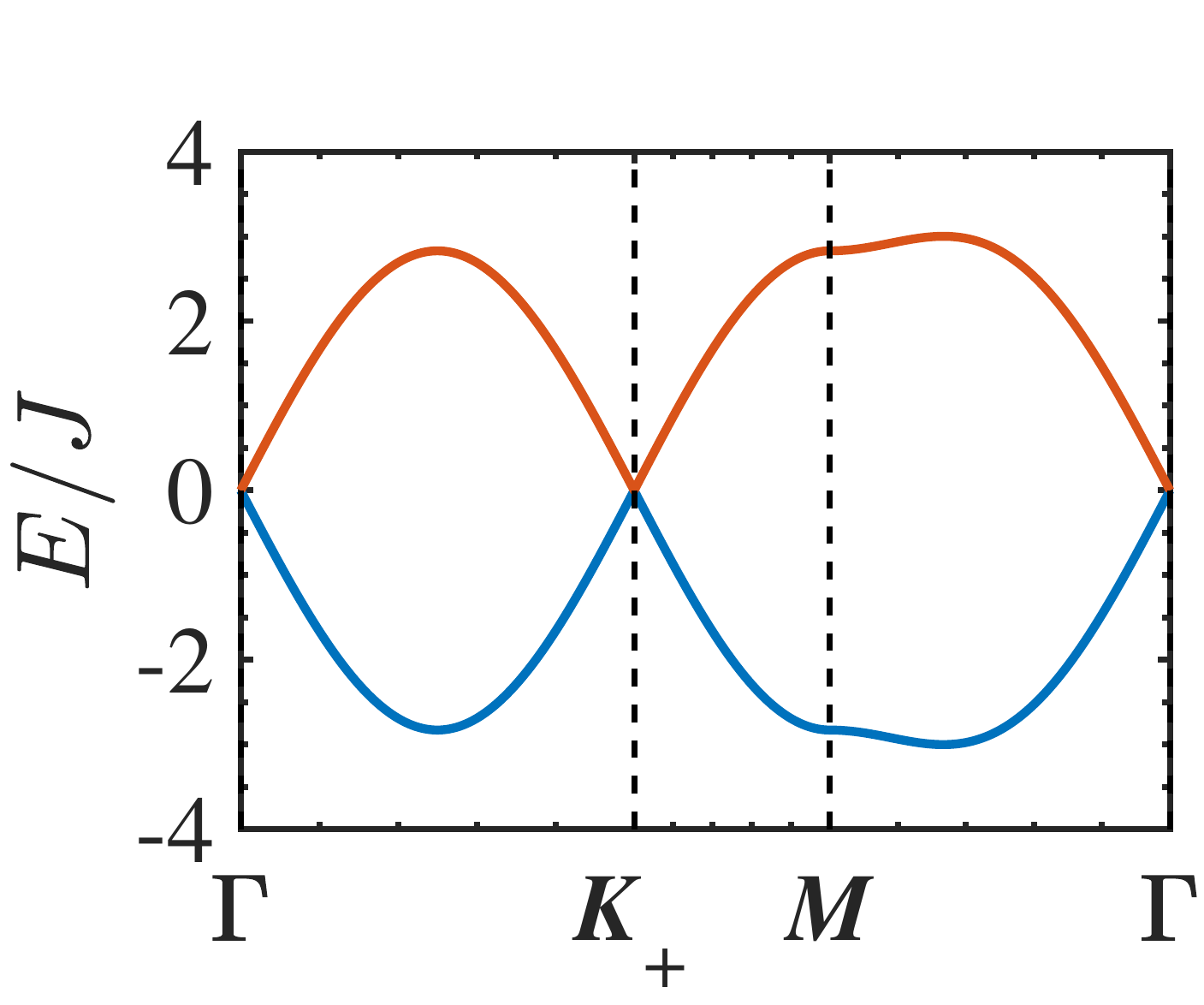}}
\end{minipage}%
\caption{(a) Schematic representation of the setup for the Mori\'e superlattices with an external spatially dynamic modulated field and a linear tilted potential. (b) Blue dots with the lower orange boundary expressing the commensurate TBG parameter space in terms of MS lattice constant $a_\mathcal{M}$ (in units of $a_0$) versus twist angle $\theta$. (c) The numerical (solid line) and analytical (red dots) prediction of the number of commensurate cases $\mathcal{N}$ versus $a_\mathcal{M}/a_0$. (d) The corresponding effective tight binding lattice with tunneling phases 0 (solid lines), $\pi$ (black arrows), and $\phi$ (red arrows). (e)
The Brillouin zone with two Dirac points at $\textbf{\textit{K}}_\pm$ driven by the force $\textbf{\textit{F}}(\phi)$ along the parallel trajectory. (f) The energy spectrum $E$ in the unit of $J$ along the high symmetry path for $\delta=0$ and $\phi=0$.}
\end{figure*}
{\em Scheme, geometry and model.}--- We start with TBG as a modulated platform, as shown in Fig. \ref{fig:1a}. Via a rotation of two equal-length linear combinations of primitive MS vectors ${\textbf{\textit{L}}_1}=i{\textbf{\textit{a}}_1} + j{\textbf{\textit{a}}_2}$ and ${\textbf{\textit{L}}_{2}}=j{\textbf{\textit{a}}_1} + i{\textbf{\textit{a}}_2}$ ($\textbf{\textit{a}}_{1,2}$ are the lattice vectors of the monolayer graphene with lattice constant $a_0$), all the commensurate twist angles $\theta$ follow $\cos \theta  = \frac{1}{2}\frac{{{i^2} + 4ij + {j^2}}}{{{i^2} + ij + {j^2}}} $\cite{PhysRevB.85.195458}, and the  corresponding MS lattice constant reads $a_{\mathcal{M}}  =\frac{{| {i - j} |a_0}}{{2\sin ( {{\theta / 2}} )}}$. In Fig. \ref{fig:1b}, $a_{\mathcal{M}}$ shows an angle-dependent evolution with mirror-symmetry around $\theta  = {30^{\circ} }$ due to $C_6$ symmetry,  and every commensurate configuration (blue dot) makes the integers $i$, $j$ in constriction to coprime pairs. The lower boundary of the commensurate space is described by $a_{\mathcal{M}}=\frac{a_0}{{2\sin ( {{\theta /  2}})}}$, fitting the orange line in Fig. \ref{fig:1b}. One can prove that the  number of commensurate situations $\mathcal{N}$ is given by the relation

\begin{equation}\label{e1}
{\mathcal N} \approx \frac{{2\sqrt 3 \pi a_{\mathcal M}^2}}{{\zeta \left( 2 \right)a_0^2}},
\end{equation}
Here, $\zeta $ refers to the Riemann zeta function \cite{Xmanual}. The numerical result for counting the blue dots in Fig. \ref{fig:1b} and the analytical result for Eq. (\ref{e1}) show respectively as red dots and solid line in Fig. \ref{fig:1c}, which indicates that ideally there would be infinite cases where the MSs hold a triangular geometry with correspondingly an infinitely large unit cell. Benefiting from the large lattice constant, it is possible to tune the electronic properties by employing dynamic manipulations on each lattice site precisely.

Enlightened by the methodology in the optical lattices \cite{PhysRevLett.111.185302}, one needs three ingredients to generate artificial gauge fields in our proposal: (i) the underlying triangular lattice, which admits tunnelings between the Moir\'e sites \cite{PhysRevX.8.031088}; (ii) a linear tilt of energy $\Delta$ along the $y$ direction to  suppress the tunneling \cite{PhysRevB.83.205135}; (iii) a temporal modulation of the on-site energies $V_d$ to bring the tunneling  back with Peierls phases \cite{PhysRevA.86.033615}.
% \begin{multline}
   % V_d=\sum_{l=1}^3 (-1)^{l-1} \Omega_l \cos{(\bf{k}_l \bf{r})} %\cos{(\omega t+(l-1)\frac{\pi}{2})}\\ \times %\cos{(\phi_l+(l-1)\frac{\pi}{2})}
% \end{multline}
 The time averaging over the rapidly oscillating terms yields an effective Hamiltonian which is time independent  (see the Supplemental Material) \cite{Suplemental}. Because of the presence of accompanying phases in the hopping terms, the translation symmetry is broken, and we are left with a lattice  divided into two sublattices denoted $A$ and $B$ as depicted in the unit cell in Fig. \ref{fig:1d}. Moreover, we consider the effect of an inversion-symmetry breaking on-site energy detuning $+\delta$ on $A$ sites and $-\delta$ on $B$ sites to end with the effective Hamiltonian of our system, as described under the tight-binding approximation:
\begin{equation}
\begin{split}
H =& {J_1}\sum\limits_{\textbf{\textit{r}}_\ell}{(a_{\textbf{\textit{r}}_\ell} ^\dag {b_{{\textbf{\textit{r}}_\ell}+{\textbf{\textit{d}}_1}}}+a_{\textbf{\textit{r}}_\ell} ^\dag {b_{\textbf{\textit{r}}_\ell-{\textbf{\textit{d}}_1}}}+h.c.)} \\
 &+ {J_2}\sum\limits_{\textbf{\textit{r}}_\ell}  {({e^{\rm{i}\phi }}a_{\textbf{\textit{r}}_\ell} ^\dag {b_{{\textbf{\textit{r}}_\ell}  + {\textbf{\textit{d}}_2}}} + {e^{ - \rm{i}\phi }}a_{\textbf{\textit{r}}_\ell} ^\dag {b_{{\textbf{\textit{r}}_\ell}  - {\textbf{\textit{d}}_2}}} + h.c.)} \\
 &+ {J_3}\sum\limits_{\textbf{\textit{r}}_\ell}  {(a_{\textbf{\textit{r}}_\ell} ^\dag {a_{{\textbf{\textit{r}}_\ell} + {\textbf{\textit{d}}_3}}} - b_{{\textbf{\textit{r}}_\ell} + {\textbf{\textit{d}}_1}}^\dag {b_{{\textbf{\textit{r}}_\ell}+ {\textbf{\textit{d}}_2}}} + h.c.)} \\
 &+ \delta \sum\limits_{\textbf{\textit{r}}_\ell}  {(a_{\textbf{\textit{r}}_\ell} ^\dag {a_{\textbf{\textit{r}}_\ell} } - b_{{\textbf{\textit{r}}_\ell} + {\textbf{\textit{d}}_1}}^\dag {b_{{\textbf{\textit{r}}_\ell}  + {\textbf{\textit{d}}_1}}})} ,
\end{split}
\label{e2}
\end{equation}
where $J_{\eta=1,2,3}$  are the nearest-neighbour tunneling amplitudes along directions $\textbf{\textit{d}}_{\eta=1,2,3}$ with additional phases shown in Fig. \ref{fig:1d}, the lattice spacing is $a_\mathcal{M}$ and, for simplicity, we assume $a_\mathcal{M}=1$ and $J_{\eta=1,2,3}=J$ in the following; $\textbf{\textit{d}}_{1}=(1,0)$ and $\textbf{\textit{d}}_{2,3}=(1/2,\pm\sqrt{3}/2)$ are the three unit vectors connecting the nearest-neighboring sites, and we specify the lattice sites by $\textbf{\textit{r}}_\ell \equiv (m,n)$; $a_{\textbf{\textit{r}}_\ell}^\dag, a_{\textbf{\textit{r}}_\ell}$ $(b_{{\textbf{\textit{r}}_\ell} + {\textbf{\textit{d}}_1}}^\dag, b_{{\textbf{\textit{r}}_\ell}+ {\textbf{\textit{d}}_1}})$ denote the creation and annihilation operators on a site belonging to the sublattice $A (B)$.

The Hamiltonian (\ref{e2}) enjoys a higher translational symmetry than that defined by $\textbf{\textit{d}}_\eta$:
\begin{equation}
\hat T_{\textbf{\textit{d}}_1} = {({e^{\rm{i}\pi }})^{{n }}}{\sigma _x}\mathcal{K}{T_{\textbf{\textit{d}}_1} },
\end{equation}
$T_{\textbf{\textit{d}}_{1,3}}$ are the translation operators that moves the lattice along the $\textbf{\textit{d}}_{1,3}$ direction by a unit vector, $\mathcal{K}$ is the complex conjugate operator, $\sigma_x$ is the Pauli matrix representing the sublattice exchange, $({e^{\rm{i}\pi }})^{{n }}$ is a  $U(1)$ local gauge transformation and $n$ is the coordinate of lattice sites in $\textbf{\textit{d}}_{3}$ direction.
Therefore, we can classify the higher translation operators denoted as ${{\hat T}_{2\textbf{\textit{d}}_1 }} = \hat T_{\textbf{\textit{d}}_1} ^2$ and ${{\hat T}_{\textbf{\textit{d}}_3} } = {T_{\textbf{\textit{d}}_3} }$, and the system obeys (in the Landau gauge, see below)
\begin{equation}
{{\hat T}_{2\textbf{\textit{d}}_1 }}{{\hat T}_{\textbf{\textit{d}}_3} }\hat T_{2\textbf{\textit{d}}_1 }^{ - 1}\hat T_{\textbf{\textit{d}}_3} ^{ - 1} = 1,
\end{equation}
  Thus \cite{Note}, we can perform a Fourier transformation to the annihilation operators as ${a_\textbf{\textit{k}}} = \frac{1}{{\sqrt N }}\sum\nolimits_{\textbf{\textit{r}}_\ell}  {{a_{\textbf{\textit{r}}_\ell}}{e^{ - \rm{i}\textbf{\textit{k}} \cdot \textbf{\textit{r}}_\ell }}} $, ${b_\textbf{\textit{k}}} = \frac{1}{{\sqrt N }}\sum\nolimits_{\textbf{\textit{r}}_\ell}  {{b_{\textbf{\textit{r}}_\ell} }{e^{ - \rm{i}\textbf{\textit{k}} \cdot \textbf{\textit{r}}_\ell}}}$ with momentum $\textbf{\textit{k}}=(k_x,k_y)$, and define a basis of two-component spinor operator as ${\psi _\textbf{\textit{k}}} \equiv {[{a_\textbf{\textit{k}}},{b_\textbf{\textit{k}}}]^T}$. The Hamiltonian (\ref{e2}) can be rewritten as $H = \sum\nolimits_{\textbf{\textit{k}}} {\psi _\textbf{\textit{k}}^\dag \mathcal{H}(\textbf{\textit{k}}){\psi _\textbf{\textit{k}}}}$, and the Bloch matrix reads
\begin{equation}\label{h4}
\mathcal{H}(\textbf{\textit{k}}) =\textbf{\textit{h}}(\textbf{\textit{k}}) \cdot \boldsymbol{\sigma},
\end{equation}
 with the Pauli matrices $ \boldsymbol{\sigma}   = ({\sigma _x},{\sigma _y},{\sigma _z})$ and the vector field in momentum space $\textbf{\textit{h}}(\textbf{\textit{k}})= ({h_x},{h_y},{h_z})$, where ${h_x} = J [1 +\cos (2{\textbf{\textit{d}}_1}\cdot\textbf{\textit{k}}+ \phi ) - \cos ({\textbf{\textit{d}}_1}\cdot\textbf{\textit{k}} + {\textbf{\textit{d}}_2}\cdot\textbf{\textit{k}} + \phi ) + \cos ({\textbf{\textit{d}}_3}\cdot\textbf{\textit{k}})]$, ${h_y} = J[\sin (2{\textbf{\textit{d}}_1}\cdot\textbf{\textit{k}}+ \phi )+ \sin ({\textbf{\textit{d}}_1}\cdot\textbf{\textit{k}} + {\textbf{\textit{d}}_2}\cdot\textbf{\textit{k}} + \phi ) -\sin ({\textbf{\textit{d}}_3}\cdot\textbf{\textit{k}})]$,
${h_z}= \delta + 2J\cos (2{\textbf{\textit{d}}_1}\cdot\textbf{\textit{k}})$.

\begin{figure}[t]
\begin{minipage}{0.5\linewidth}
\subfigure{\label{fig:2a}\subfigimg[width=4.2cm,height=4.2cm]{(a)}{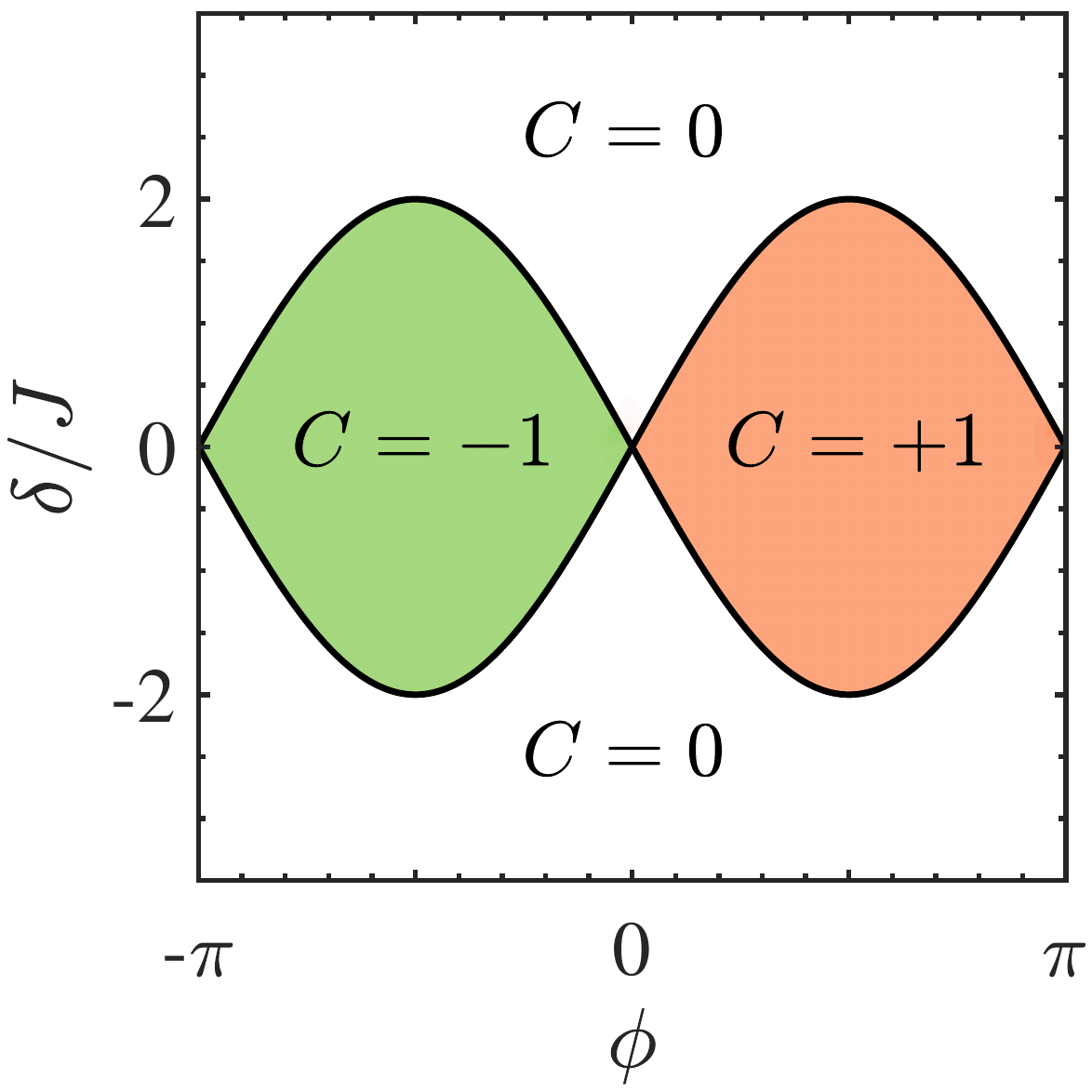}}
\end{minipage}%
\begin{minipage}{0.5\linewidth}
\subfigure{\label{fig:2b}\subfigimg[width=4.3cm,height=4.3cm]{(b)}{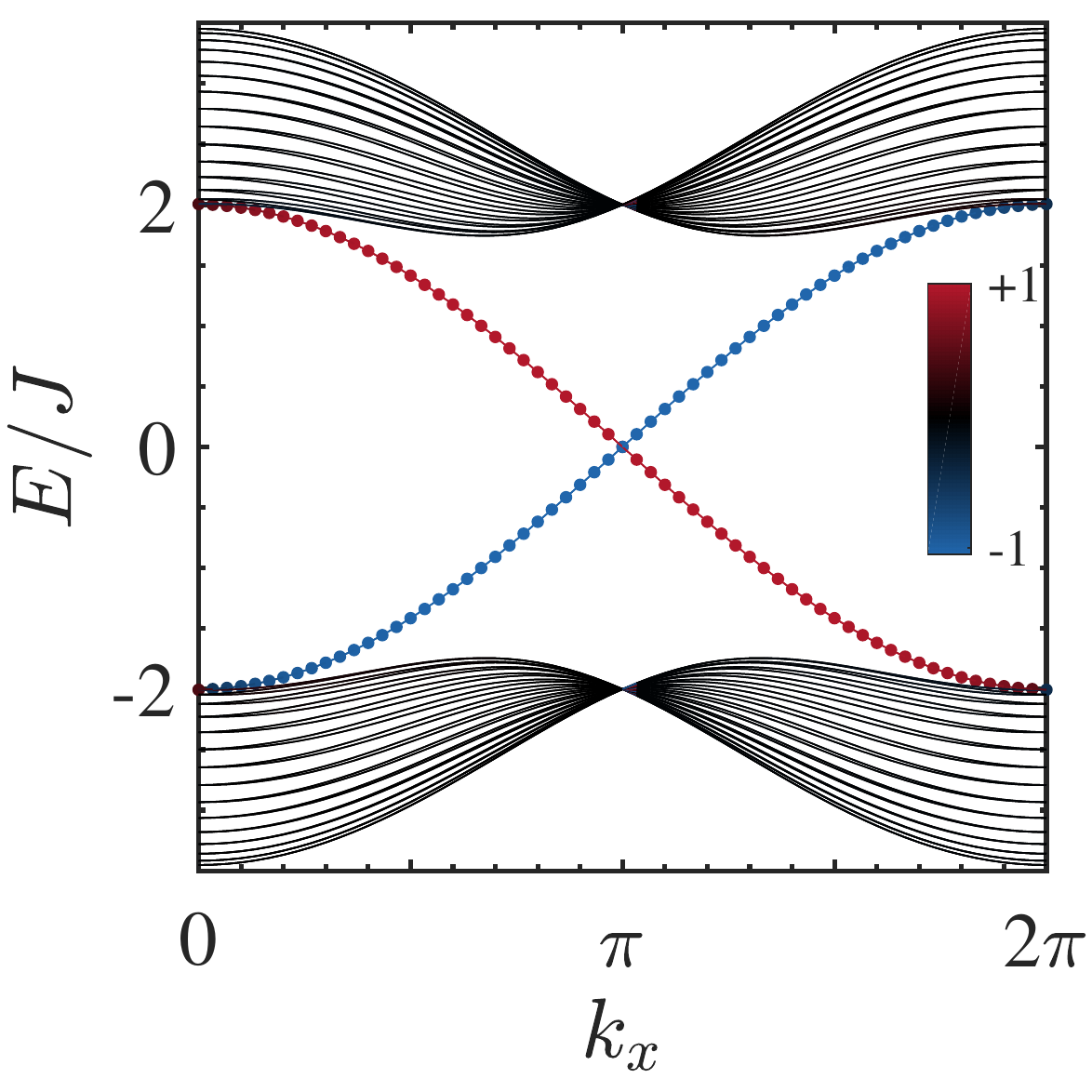}}
\end{minipage}%
\caption{(a) Phase diagram of the model in the $\delta-\phi$ plane with boundaries $\delta=\pm 2J\sin\phi$, and topological phase correspond to Chern numbers $C = \pm 1, 0$, respectively. (b) Band spectrums for a ribbon of the Mori\'e superlattices showing the bulk (black) and counterpropagating edge states (blue and red) inside the gaps, for $\phi=\pi/2$ and $\delta=0$.}
\end{figure}

By diagonalizing Eq. (4), we can obtain the dispersion relation as $E(\textbf{\textit{k}}) =  \pm |\textbf{\textit{h}}|=\pm \sqrt {h_x^2 + h_y^2 + h_z^2} $. For $\phi=0$ and $\delta=0$, the band structure along the symmetry path in the mini-Brillouin zone (mBZ) of the Moir\'e superlattice  [Fig. \ref{fig:1e}] is presented in Fig. \ref{fig:1f}, and the degenerate points pin to high-symmetry momenta $\textbf{\textit{K}}_{\pm}$ and $\boldsymbol{\Gamma}$ due to the threefold rotational symmetry. At half filling, the low-energy physics of the system is dominated by the quasiparticles behaving like massless relativistic Dirac fermions at two valleys ${\textbf{\textit{K}}_{\pm}}= ( {\frac{\pi }{2},\frac{{\sqrt 3 }}{6}\pi  \mp \frac{{\sqrt 3 }}{3}\pi } )$. Distinguished from the Haldane model with a mass related to momentum $\textbf{\textit{k}}$ \cite{PhysRevLett.61.2015}, the effective Dirac fermions here acquire a mass that depends on both on $\textbf{\textit{k}}$ and $\phi$, and their momentum space trajectories are driven by the force $\textbf{\textit{F}}(\phi ) = ({F_x}(\phi ),{F_y}(\phi ))$  \cite{PhysRevLett.112.155302,PhysRevLett.121.033904}. The Dirac fermions here can be treated as experiencing a \textit{phase-dependent} gauge potential, which leads to the substitution $({k_x},{k_y}) \to ({k_x} + {F_x}(\phi ),{k_y} + {F_y}(\phi ))$ in the Bloch Hamiltonian. As the Dirac cones are still present, their positions in quasi-momentum space are shifted by  $\textbf{\textit{F}}= (0,\frac{{2\sqrt 3 }}{3}\phi )$. As a result, the Hamiltonian (\ref{h4}) can be linearized as
\begin{equation}\label{h5}
\begin{split}
H(\textbf{\textit{p}}) =& [\delta \mp J(2\sin{\phi}  + \cos{\phi} p_x - \sqrt 3 \cos{\phi} {p_y})]{\sigma _z}\\
 & \pm J({p_x} + \sqrt 3 {p_y}){\sigma _x} + 2J{p_x}{\sigma _y},
\end{split}
\end{equation}
where $\textbf{\textit{p}} = \textbf{\textit{k}} + \textbf{\textit{F}}(\phi ) - {\textbf{\textit{K}}_{\pm} }$. From Hamiltonian (\ref{h5}), we can get a Haldane mass term $m_ { \pm }  =\delta \mp  2J\sin \phi$. The vanishing mass defines the boundary between the topologically non-trivial Chern insulator and trivial band insulator in the phase diagram as shown in Fig. \ref{fig:2a}. The Chern number is given as
\begin{equation}
C = \frac{1}{2}\left[ {{\rm{sgn}}\left( {\delta {\rm{ + }}2J\sin \phi } \right) - {\rm{sgn}}\left( {\delta  - 2J\sin \phi } \right)} \right].
\end{equation}

A general understanding of the topological properties of the system requires to determine its corresponding topological invariant. In particular, the lowest energy band of the Chern number $C$ can be calculated using the following expression
\begin{equation}
C = \frac{1}{{4\pi }}\iint_{\rm BZ} {\frac{\textbf{\textit{h}}}{{{{\left| \textbf{\textit{h}} \right|}^3}}} \cdot \left( {\frac{{\partial \textbf{\textit{h}}}}{{\partial {k_x}}} \times \frac{{\partial \textbf{\textit{h}}}}{{\partial {k_x}}}} \right)d{k_x}d{k_y}}.
\end{equation}
The phase diagram of $C$ as a function of $\delta/J$ and $\phi$ is shown in Fig. \ref{fig:2a}, which depicts a novel Haldane-like phase. In fact, the topological invariant here has a simple geometric interpretation: the Chern number of a two-band model is equivalent to the winding number of the mapping from a two dimensional (2D) Brillouin zone (BZ)  which is 2-torus $\mathbb{T}^2$ to the 2-sphere $\mathbb{S}^2$ (ie., Bloch sphere) \cite{PhysRevB.74.045125}.

 According to the phase diagram, we have a topologically non-trivial phase if $|\delta| < 2J\sin \phi$. As a consequence of the bulk-edge correspondence, one obtains chiral edge modes whenever the Fermi energy resides in the band gap. To demonstrate this, we use the same tight-binding approach to model a quasi-one-dimensional stripe in the dashed line of Fig. \ref{fig:1d}, consisting of 30 unit cells. Figure \ref{fig:2b} shows the counterpropagating edge states in the gap, and the red (blue) color represents the degree of localization $\rho=\frac{{{{ | {{\psi _U}} |}^2} - {{ | {{\psi _D}} |}^2}}}{{{{ | {{\psi _U}} |}^2} + {{ | {{\psi _D}} |}^2}}}$ on the up (down) edges, which is calculated from the wave function $\psi_{U (D)}$ densities on the edge chains.

{\em Measuring topological quantities.}---In order to further investigate the topology of our system, we use a state tomography approach from quench dynamics, in which the Hopf number is well defined in the $(2+1)$-dimensional (i.e., 2D momentum $\textbf{\textit{k}}$ plus 1D time $\tilde t$) space \cite{PhysRevLett.118.185701}. Here we consider a sudden change of $\textbf{\textit{h}}(\textbf{\textit{k}})$ from an initial topologically trivial $\textbf{\textit{h}}_i(\textbf{\textit{k}})$ to a final topologically non-trivial $\textbf{\textit{h}}_f(\textbf{\textit{k}})$. The initial wave function $| \psi ( {\textbf{\textit{k}},\tilde t = 0} ) \rangle$ is taken as the lower-band eigenstate of the initial Hamiltonian. When $\delta$ tends to infinity, we can set $| \psi ( {\textbf{\textit{k}},\tilde t = 0} ) \rangle\equiv( {0, 1} )^T$. After the quench, the initial state evolves by final Hamiltonian $\mathcal{H}_f={\textbf{\textit{h}}_f}(\textbf{\textit{k}})\cdot\boldsymbol{\sigma}$, and gives rise to a time-dependent state $| {\psi ( {\textbf{\textit{k}},\tilde t})} \rangle  = \exp ( { - {\rm{i}}{\textbf{\textit{h}}_f}( \textbf{\textit{k}} ) \cdot \boldsymbol{\sigma} \tilde t} )| {\psi ( {\textbf{\textit{k}},\tilde t = 0} )} \rangle$.  From this state, we can define a Bloch vector which lies on $\mathbb{S}^2$ as $\textbf{\textit{n}}( {\textbf{\textit{k}},\tilde t} ) = \langle {\psi ( {\textbf{\textit{k}},\tilde t} )} |\boldsymbol{\sigma} | {\psi ( {\textbf{\textit{k}},\tilde t} )} \rangle$. The three components of $\textbf{\textit{n}}\left( {\textbf{\textit{k}},\tilde t} \right)$  are given by
\begin{figure}[t]
\begin{minipage}{1\linewidth}
\subfigure{\label{fig:3a}\subfigimgt[width=8.3cm,height=1.92cm]{(a)}{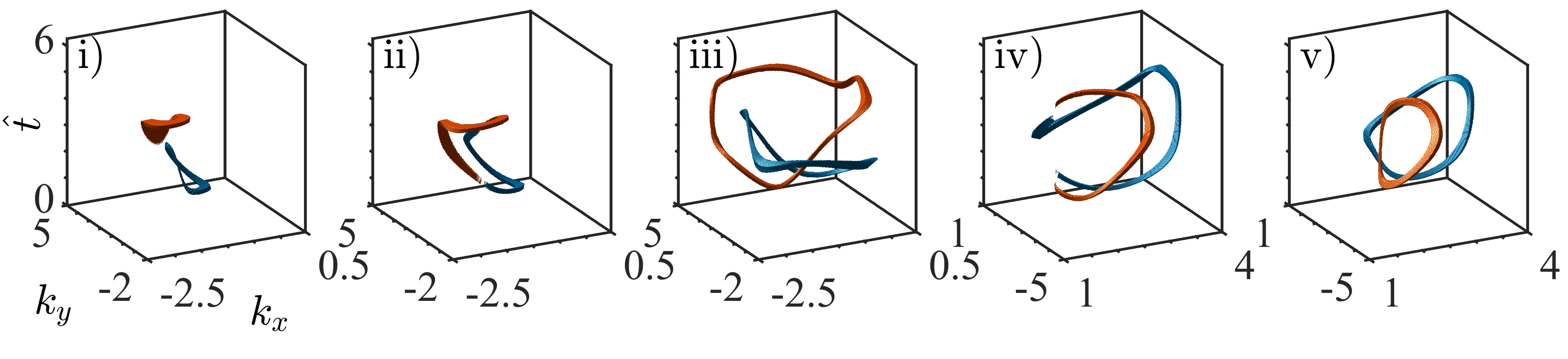}}
\end{minipage}\\
\vspace{-2mm}
\begin{minipage}{1\linewidth}
\subfigure{\label{fig:3b}\subfigimgt[width=8.3cm,height=1.92cm]{(b)}{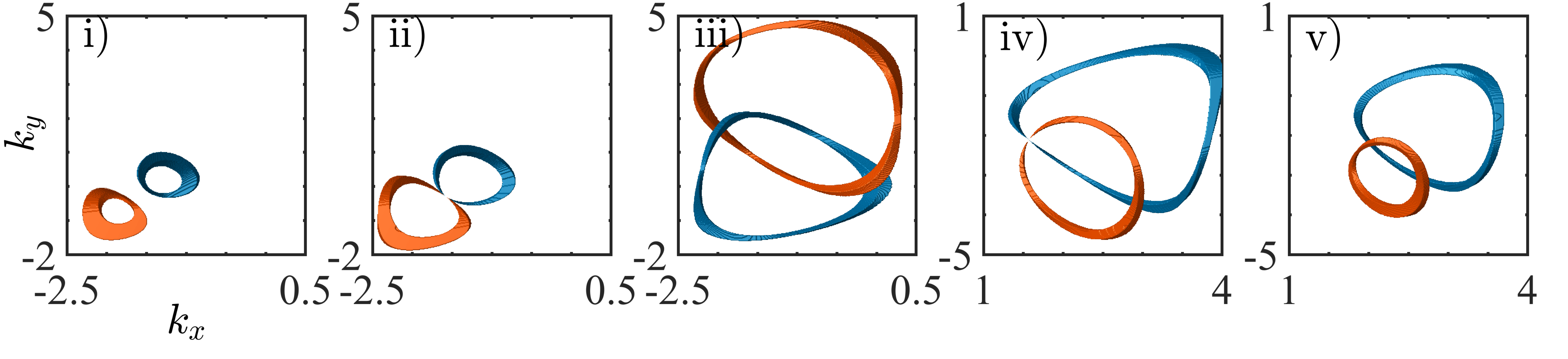}}
\end{minipage}\\
\vspace{-4mm}
\begin{minipage}{1\linewidth}
\subfigure{\label{fig:3c}\subfigimg[width=6.3cm,height=2.2cm]{(c)}{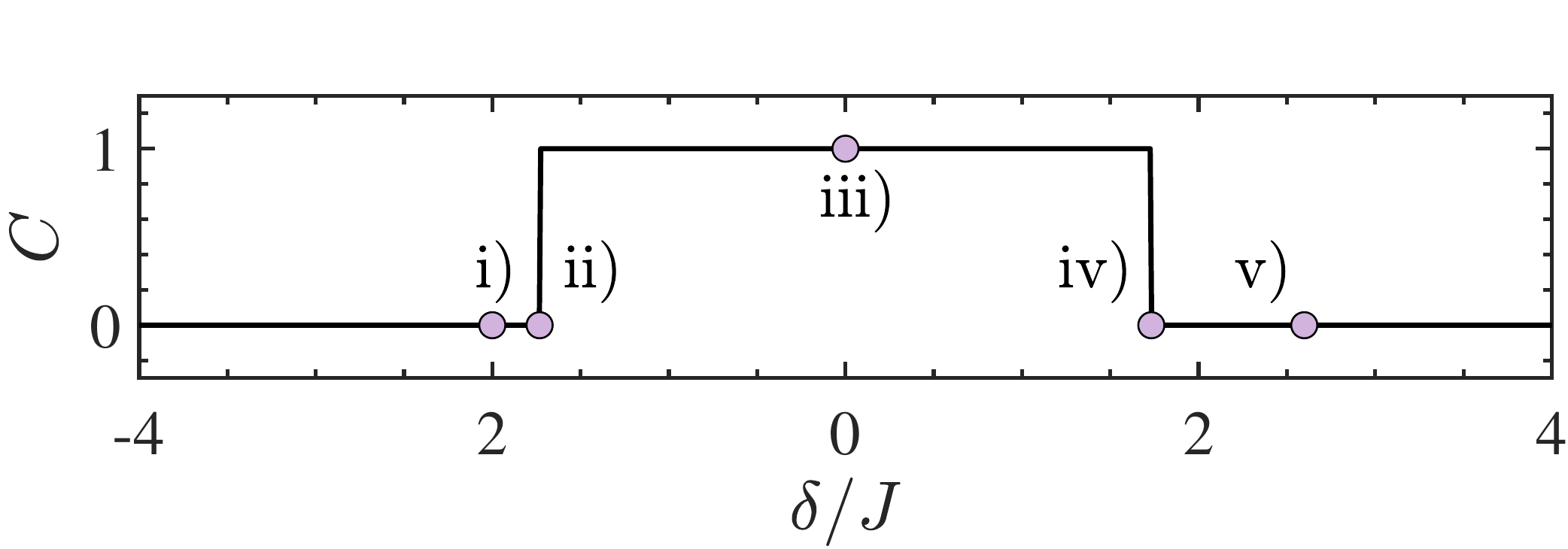}}
\subfigure{\label{fig:3d}\subfigimg[width=2.2cm,height=2.2cm]{(d)}{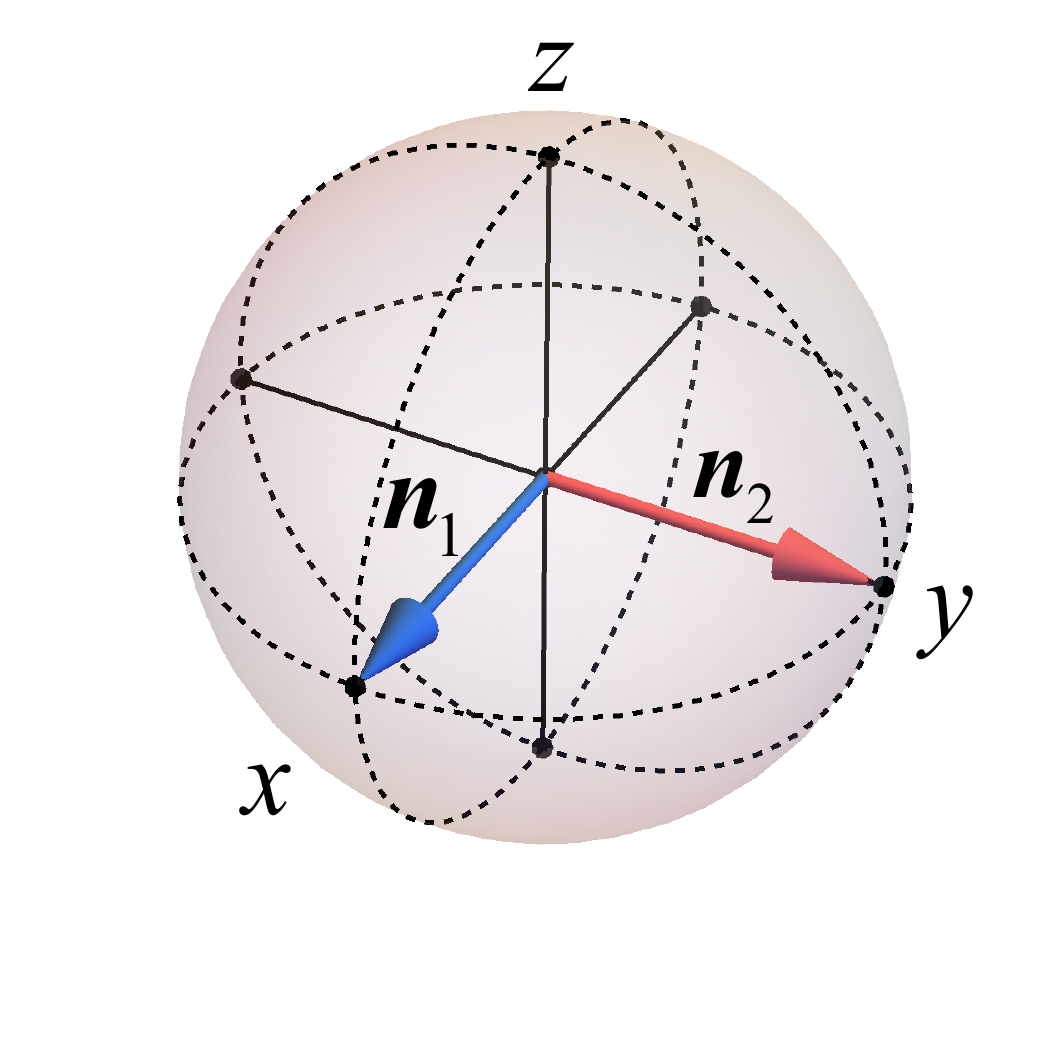}}
\end{minipage}\\
\vspace{-2mm}
\begin{minipage}{1\linewidth}
\subfigure{\label{fig:3e}\subfigimgt[width=8.3cm,height=1.92cm]{(e)}{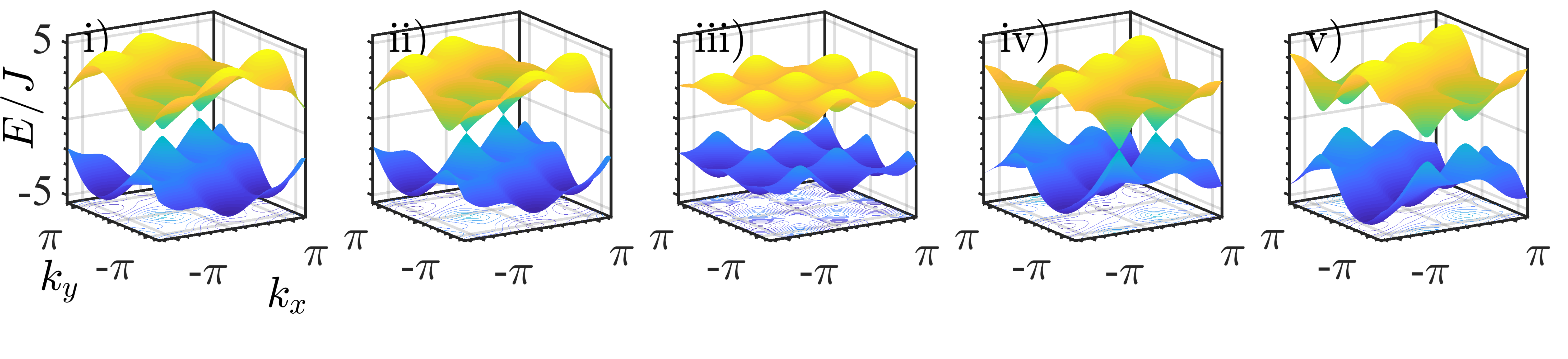}}
\end{minipage}%
\caption{Mapping out the  loci of topological phase transition versus the detuning $\delta$. Here we set fixed $\phi=\pi/3$, and from i) to iv), $\delta/ J=-2, -\sqrt{3}, 0, \sqrt{3}, 3\sqrt{3}/2$, respectively. (a) Topological links between the
pre-images in the momentum space $(k_x,k_y,\hat{t})$  from vector $\textbf{\textit{n}}_{1,2}$  on the Bloch sphere (d). (b) The contours of links  presented as a top view of (a). (c) The Chern number obtained from the linking number of these contours (or the absence of a contour) and plotted versus $\delta$. (d) The Bloch sphere with vectors $\textbf{\textit{n}}_1=(1,0,0)$ and $\textbf{\textit{n}}_2=(0,1,0)$.  (e) Calculated band structures for various detunings to illustrate the opening and closing of the Dirac points through the topological phase transitions.}
\end{figure}
\begin{equation}
\begin{split}
{n_x} =& \sin (\hat t){{\hat h}_y} + [1 - \cos (\hat t)]{{\hat h}_x}{{\hat h}_z},\\
{n_y} =&  - \sin (\hat t){{\hat h}_x} + [1 - \cos (\hat t)]{{\hat h}_y}{{\hat h}_z},\\
{n_z} =& \cos (\hat t) + [1 - \cos (\hat t)]\hat h_z^2,
\end{split}
\end{equation}
where $\textbf{\textit{h}}_f / |\textbf{\textit{h}}_f|=({{{\hat h}_x},{{\hat h}_y},{{\hat h}_z}})$ is the normalized post-quench Hamiltonian vector, and $\hat t = 2|{{\textbf{\textit{h}}_f}}|\tilde t \in \left[ {0,2\pi } \right) \cong {\mathbb{S}^1}$ is the rescaled time, serving as a periodicity in the time direction. The Hopf map $f:{\mathbb{T}^3} = {\mathbb{T}^2} \times {\mathbb{S}^1} \to {\mathbb{S}^2}$ illustrates a projection from the momentum space described by $({{k_x},{k_y},\hat t}) \in {\mathbb{T}^3}$ to ${\mathbb{S}^2}$. For the Hopf map, the pre-image $f^{-1}(\textbf{\textit{n}})$ of each point on the ${\mathbb{S}^2}$ corresponds to a closed loop in the torus ${\mathbb{T}^3}$, and all these loops are topologically linked to each other, forming the Hopf fibration. Here, the linking number (Hopf invariant) of two pre-images $f^{ - 1}$ topologically equals the Chern number of the post quench Hamiltonian $\mathcal{H}_f$ and this number is invariant under continuous deformation of $f$.

For instance, we take two constant vectors $\textbf{\textit{n}}_{1,2}$ on the Bloch sphere [Fig. \ref{fig:3d}],
and Fig. \ref{fig:3a} shows two trajectories of $f^{-1}(\textbf{\textit{n}}_{1,2})$ in the $(k_x,k_y,\hat{t})$ space with respect to the different detuning $\delta$. From the top view of Fig. \ref{fig:3a}, we can easily count the linking number of the contours in the Fig. \ref{fig:3b}  and thereby obtain the Chern number of the final Hamiltonian $\mathcal{H}_f$. By continuously changing $\delta$, we obtain the phase diagram, shown in Fig. \ref{fig:3c}, which features a topologically non-trivial interval of $\delta$ with Chern number $C=1$ between topologically trivial regions with Chern number $C=0$.
The Chern number measured by the quench dynamics approach agrees well with the theoretical prediction obtained from Fig. \ref{fig:2a}, and the corresponding band structures in Fig. \ref{fig:3e} also fit the opening and closing behaviour through the topological phase transitions.

{\em Conclusion.}---Summing up, we introduce and analyze a novel scheme to create tunable, artificial gauge fields in the periodically modulated Moir\'e superlattices. By imprinting Peierls phase on the hopping parameters, we generalize a Haldane-like model and investigate its topological properties. Moreover, we introduce a quench dynamic proposal to directly quantify the topological order in this model. As a fully controllable system,  periodically modulated Moir\'e superlattices will provide an ideal platform to simulate and realize these topological models, and widen research perspectives in novel topological phases of matter.
\bigbreak
The authors acknowledge financial support from the
King Abdullah University of Science and Technology
(KAUST).
%\cite{berry1984quantal} \cite{PhysRevB.81.161405}

%We can rewiten
%This can be rewritten into a second quantized, time-dependent Hamiltonian of the
%form
%\begin{thebibliography}{1}
%
%\bibitem{DSGolubevPRB2001}
%D.~S. Golubev and A.~D. Zaikin,
%\newblock Phys. Rev. B {\bf 64}, 014504 (2001).
%
%\bibitem{DMeidanPhysicaC2008}
%D.~Meidan, Y.~Oreg, G.~Refael, and R.~A. Smith,
%\newblock Physica C {\bf 468}, 341 (2008).
%
%\bibitem{JMKosterlitzJPHYSC1973}
%J.~M. Kosterlitz and D.~J. Thouless,
%\newblock J. Phys. C {\bf 6}, 1181 (1973).
%
%\end{thebibliography}
%
\end{document}